\documentclass{emulateapj}
\usepackage{graphicx,color}
\usepackage{amssymb,amsmath}
\usepackage[colorlinks,hyperfootnotes=false,citecolor=blue,linkcolor=blue]{hyperref}

\begin{document}
\shorttitle{SL Analysis of MACS2135}
\shortauthors{Zitrin \& Broadhurst}

\slugcomment{Submitted to the Astrophysical Journal Letters}

\title{Strong Lensing Analysis of the Powerful Lensing Cluster MACS J2135.2-0102 ($\lowercase{z}$=0.33)}


\author{Adi Zitrin\altaffilmark{1,2} \& Tom Broadhurst\altaffilmark{3,4}} 
\altaffiltext{1}{Cahill Center for Astronomy and Astrophysics, California Institute of Technology, MC 249-17, Pasadena, CA 91125, USA; adizitrin@gmail.com}
\altaffiltext{2}{Hubble Fellow}
\altaffiltext{3}{Department of Theoretical Physics, University of Basque Country UPV/EHU, Bilbao, Spain}
\altaffiltext{4}{IKERBASQUE, Basque Foundation for Science, Bilbao, Spain}


\begin{abstract}
We present a light-traces-mass (LTM) strong-lensing model of the massive lensing cluster MACS J2135.2-0102 ($z$=0.33; hereafter MACS2135), known in part for hosting the Cosmic Eye galaxy lens. MACS2135 is also known to multiply-lens a $z=$2.3 sub-mm galaxy near the Brightest Cluster Galaxy (BCG), as well as a prominent, triply-imaged system at a large radius of $\sim$37\arcsec\ south of the BCG. We use the latest available Hubble imaging to construct an accurate lensing model for this cluster, identifying six new multiply-imaged systems with the guidance of our LTM method, so that we have roughly quadrupled the number of lensing constraints. We determine that MACS2135 is amongst the top lensing clusters known, comparable in size to the Hubble Frontier Fields. For a source at $z_{s}=2.32$, we find an effective Einstein radius of $\theta_{e}=27\pm3 \arcsec$, enclosing $1.12 \pm0.16 \times10^{14}$ $M_{\odot}$.  We make our lens model, including mass and magnification maps, publicly available\footnote{ftp://wise-ftp.tau.ac.il/pub/adiz/MACS2135/}, in anticipation of searches for high-$z$ galaxies with the {\it James Webb Space Telescope} for which this cluster is a compelling target. \vspace{0.05cm}
\end{abstract}


\section{Introduction}\label{intro}

Strong gravitational lensing (SL) by galaxy clusters has by now become a reliable, routine tool in Astronomy. Multiply-imaged background galaxies allow us to map in detail the otherwise-invisible dark matter (DM) distribution of the cluster, as well as to detect faint background objects that are highly magnified by the foreground cluster lens \citep[see reviews by][]{Kneib2011review,Bartelmann2010reviewB}.

The past decade in particular has seen a dramatic increase in SL-related science, thanks mainly to the continued impressive performance of the {\it Hubble Space Telescope}, from the combination of deep high-resolution optical and NIR imaging, and because of the development of improved lens modeling techniques \citep[e.g.][]{Broadhurst2005a,Diego2005Nonparam,Jullo2007Lenstool,Liesenborgs2006,Zitrin2009_cl0024}. Cluster lensing programs such as the Cluster Lensing and Supernova with Hubble (CLASH; PI: Postman, \citealt{PostmanCLASHoverview}), and the ongoing Hubble Frontier Fields (HFF; PI: Mountain \& Lotz; see \citealt{Lotz2016HFF}) with HST, have proven extremely successful for SL, including the detection of {\it hundreds} of multiply lensed \citep[e.g.][as few examples]{Monna2014RXC2248,Jauzac2014M0416,Zitrin2014CLASH25} and high-redshift, magnified background objects extending into the reionization era above $z\gtrsim6$   \citep{Bradley2013highz,Atek2015HalfHFF_LF,Coe2014FF,Zheng2012NaturZ}, and beyond, to the current limits of detection at z$\sim11$ (\citealt{Coe2012highz,Zitrin2014highz}). Construction of luminosity functions is feasible now to $z\sim9$ (\citealt{Atek2015HalfHFF_LF,Mcleod2016,Livermore2016LF}). Several lensed supernova have been discovered \citep[e.g.][]{Patel2014SN} including the first multiply-imaged supernova as a quadrupole Einstein cross, and its subsequent reappearance \citep{Kelly2016reappearance}. Detailed studies of large highly magnified galaxies at $z\sim1-5$ have helped constrain UV-escape fractions below the Ly-limit \citep[e.g.][]{Nicha2016escape}, metallicity gradients and outflows \citep{Tucker2015} and star-formation details \citep[e.g.][]{Wuyts2012}. Cosmological models have been examined with SL through arc and Einstein radius statistics \citep{OguriBlandford2009,Horesh2010,Waizmann2012JeanClaude0717} and multi-wavelength related discoveries have been made of magnified, X-ray, radio or sub-mm galaxies \citep[e.g.][]{van_Weeren2016,Gonzales2016}.

\begin{figure*}
 \begin{center}
   \includegraphics[width=157mm,trim=0cm 0cm 0cm 0cm,clip]{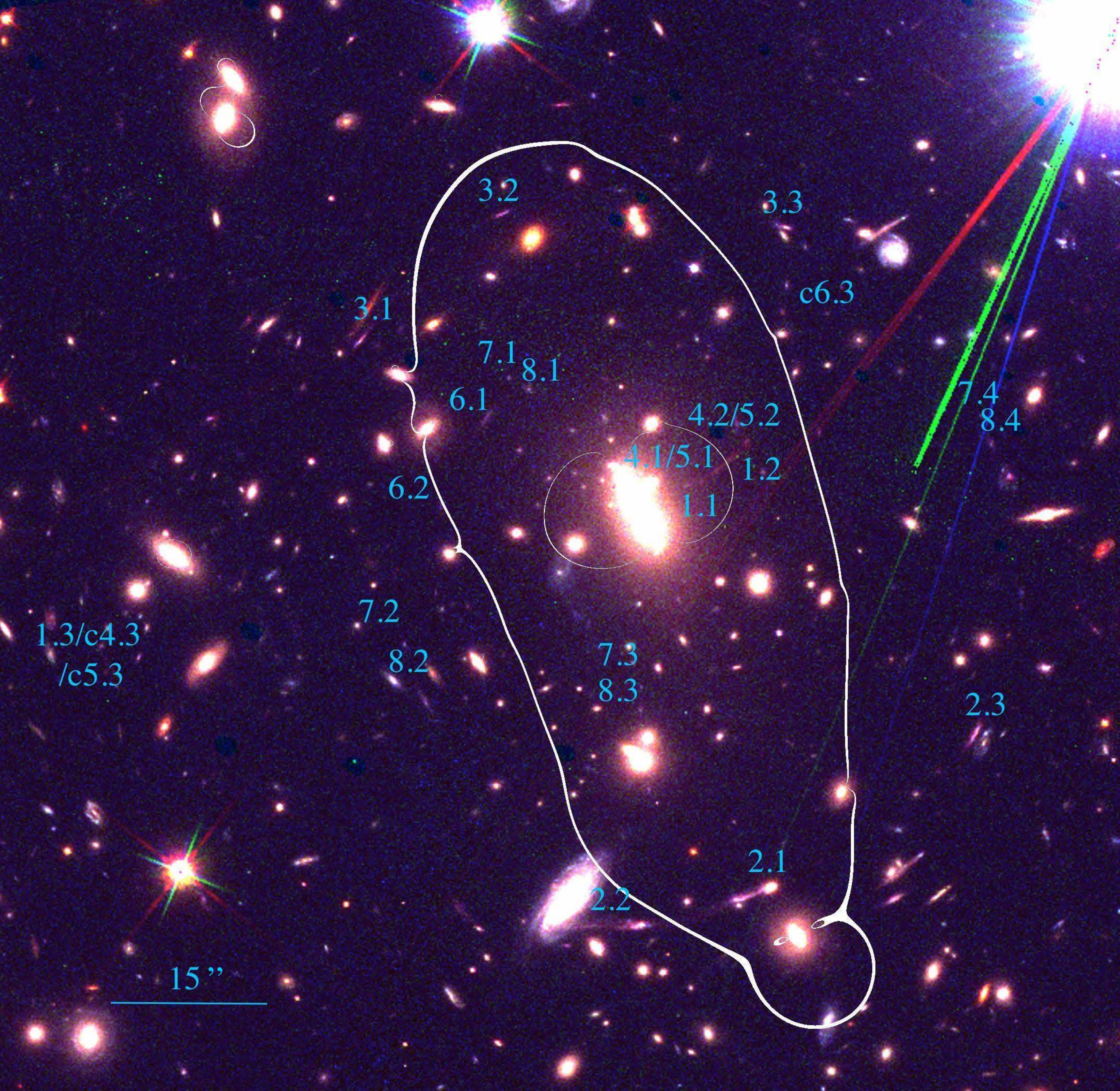}
 \end{center}
\caption{The central field of the galaxy cluster MACS2135 (R$=$[F140W+F110W]; G$=$F814W; B$=$F606W). Multiple-images and candidates, most of which (aside for systems 1 and 2) were found in this work, are indicated, and the resulting critical curves from our model are overlaid for $z_{s}=2.32$, revealing the large size and relatively high ellipticity of this lensing cluster (critical curve major-to-minor axis ratio of $\sim2.5$).}\vspace{0.5cm}
\label{fig1}
\end{figure*}

This progress in SL is inspiring new campaigns including the reionization cluster survey, RELICS (PI: Coe), informed by the CLASH and HFF programs dedicated to SL, and designed to enhance lensing-enabled science with future facilities and in particular, the {\it James Webb Space Telescope} (JWST). Aside from the immediate science goals, part of the underlying motivation in these programs is to discover and characterize the ``best'' lensing targets for JWST for optimizing the detection of very distant background objects that lie beyond the reach of \emph{Hubble}. Since there are many massive clusters in the sky \citep[e.g.][]{OguriBlandford2009,Waizmann2012JeanClaude0717}, choosing the largest and most powerful lenses requires systematic lens modeling of controlled samples of clusters with continued space imaging for the detection of the multiply lensed images required for this purpose. We are also using the HST archive for progressing in this work with backlog of numerous unanalyzed massive clusters, including the data analyzed here as well as other X-ray selected clusters from the MAssive Cluster Survey (MACS; \citealt{Ebeling2010FinalMACS}). 

We begin our systematic analysis with MACS J2135.2-0102 ($z$=0.33; hereafter MACS2135), which exhibits several prominent arcs ranging up to $\gtrsim40\arcsec$ from the Brightest Cluster Galaxy (BCG), but lacks a recent lensing analysis that takes advantage of the achieved Hubble data. MACS2135 has been the subject of various previous studies. In particular, it became known as the cluster host of the Cosmic Eye galaxy-galaxy lens \citep{Smail2007CosmicEye,Stark2008NaturCosmicEye}, one of the most distant clear examples of a typical star-forming galaxy at $z=3.1$. MASC2135 was later found to multiply-image a prominent sub-mm galaxy \citep{Swinbank2010Natur,Ivison2010}. In their analysis, \citet{Swinbank2010Natur} constructed a SL model for this cluster, based on the sub-mm galaxy system - for which they measured a spectroscopic redshift of $z=2.3259$ and identified a third counter image on the east side of the cluster. They also used and measured a redshift for $z=2.32$  for a triply-imaged galaxy at a remarkable distance of $\sim37\arcsec$ south of the BCG, two of its images straddling the critical curve into a giant arc. We did not find records of other, recent SL models for this cluster.

Here we make use of the most recent HST imaging, that extends significantly the coverage of earlier work described above, to enhance the lens model with many new multiple-images and to make this publicly-available given the expected large critical area (the model of \citealt{Swinbank2010Natur} implied an Einstein radius of $\sim35\arcsec$) and relatively high ellipticity, which enhances the cross section of lensing clusters \citep[][and references therein]{Zitrin2013M0416}. The paper is organized as follows. We present the observations in \S \ref{obs}, and the SL modeling in \S \ref{lensmodel}. We conclude the work and discuss the results in \S \ref{discussion}. Throughout we use a standard $\Lambda$CDM cosmology with $\Omega_{\rm m0}=0.3$, $\Omega_{\Lambda 0}=0.7$, $H_{0}=100$ $h$ km s$^{-1}$Mpc$^{-1}$, $h=0.7$, and magnitudes are given using the AB convention. 1\arcsec\ equals 4.75 kpc at the redshift of the cluster. Errors are $1\sigma$ unless otherwise stated.

\vspace{0.5cm} 

\section{Data and Observations} \label{obs}

The HST archive lists two ``sets'' of imaging for MACS2135 - one targets the Cosmic Eye galaxy-galaxy lens system in the northern part of the cluster, and the other one targets the cluster core itself. We use here the set targeting the cluster, obtained with the \emph{Advanced Camera for Surveys} (ACS) and WFC3 cameras. Unlike the previous data available from WFPC2 and NIC2, these data cover the full area of interest and with finer resolution and sensitivity needed for the identification of more multiple images. The data we use here includes imaging in two bands with the ACS: F606W, total exposure time of 1200s, taken on 2006-05-08 (program ID 1049, PI: Ebeling); and F814W, with a total exposure time of 1440s, taken on 2013-08-19 (program ID: 12884, PI: Ebeling); and two bands with the WFC3/IR: F110W and F140W, for a total of 705.88s each, taken on 2011-08-23 (program ID 12166, PI: Ebeling). Reduced data was obtained directly from the Hubble Legacy Archive\footnote{http://hla.stsci.edu/hlaview.html}.

We ran SExtractor \citep{BertinArnouts1996Sextractor} on each of the ACS images separately, and in dual mode on the twin WFC3/IR images. We then cross-matched the outputs and generated a master photometric catalog. We ran the Bayesian Photometric Redshift program (BPZ; \citealt{Benitez2000}) on the catalog to obtain Spectral Energy Distributions (SEDs) and photometric redshifts - which are useful for enhancing the confidence in the identification of multiple images (Table \ref{multTable}). In addition, two spectroscopic redshifts, for the two systems previously known, were adopted from \citet{Swinbank2010Natur}.

We selected red-sequence cluster members, down to a F606W
magnitude of $R_{606}\leq24$ AB, from a color-magnitude diagram
made using the F814W and F606W bands. In particular we plot the $[I_{814}-R_{606}]$ color versus the $R_{606}$ magnitude, and the red sequence is immediately evident. We perform an initial selection of objects lying within 0.4 mags from the red-sequence line defined as $[I_{814}-R_{606}] = 0.0403 \times R_{606}-1.7544$. We scrutinized
by eye and slightly edited the selection to account
for objects that may have been missed, and to remove stars and several other objects that fall within the color limits but which appear on closer inspection to be in the foreground or background.  
In total our final cluster-member list includes 75 galaxies, and we supply the list online along with our mass model. The final list of cluster members and their luminosities is the starting point for our mass modeling, as we now
detail in \S \ref{lensmodel}.

\begin{figure}
 \begin{center}
   \includegraphics[width=93mm,trim=2cm -3cm 0cm 0cm,clip]{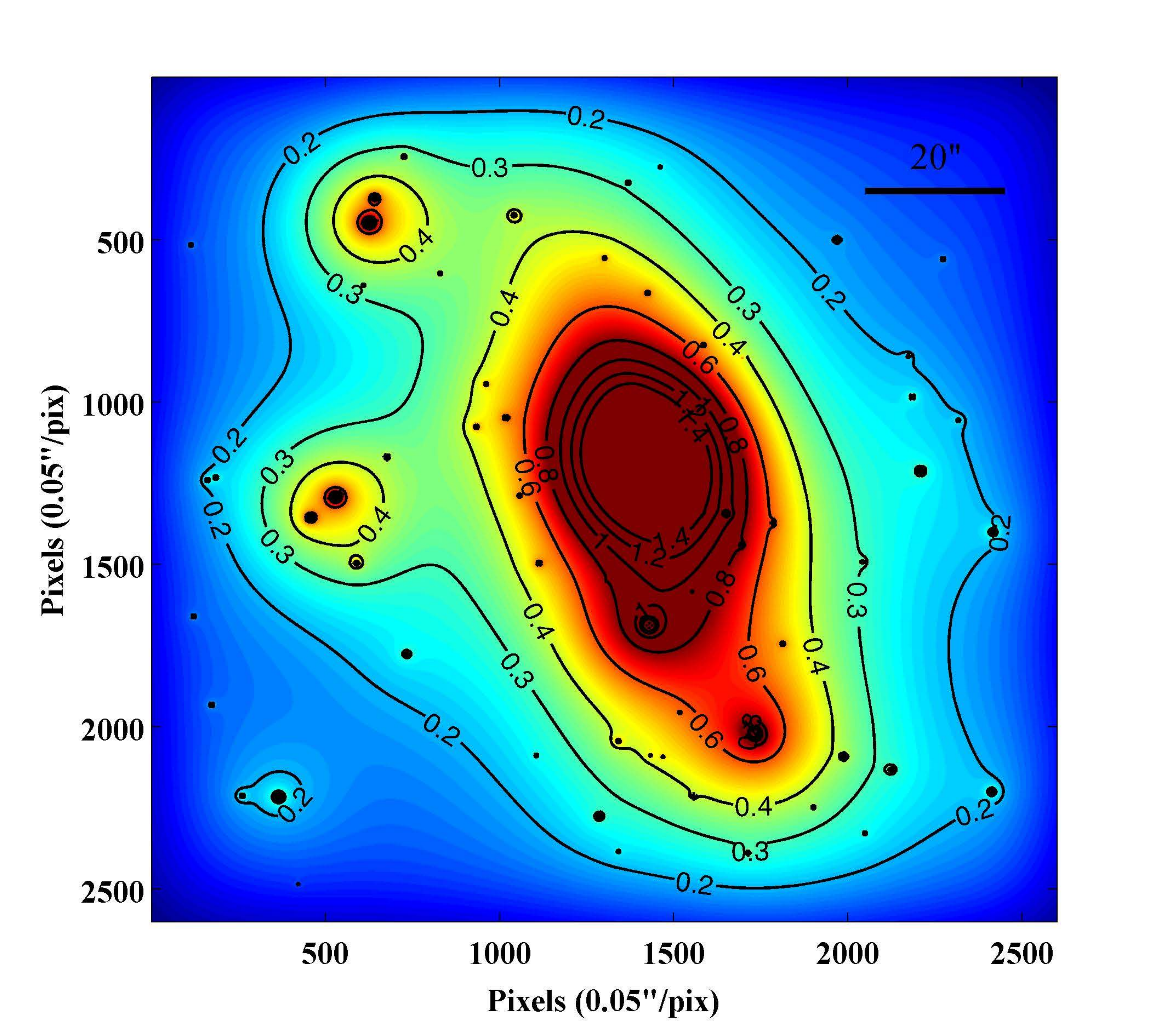}\\
   \includegraphics[width=93mm,trim=2cm 0cm 0cm 0cm,clip]{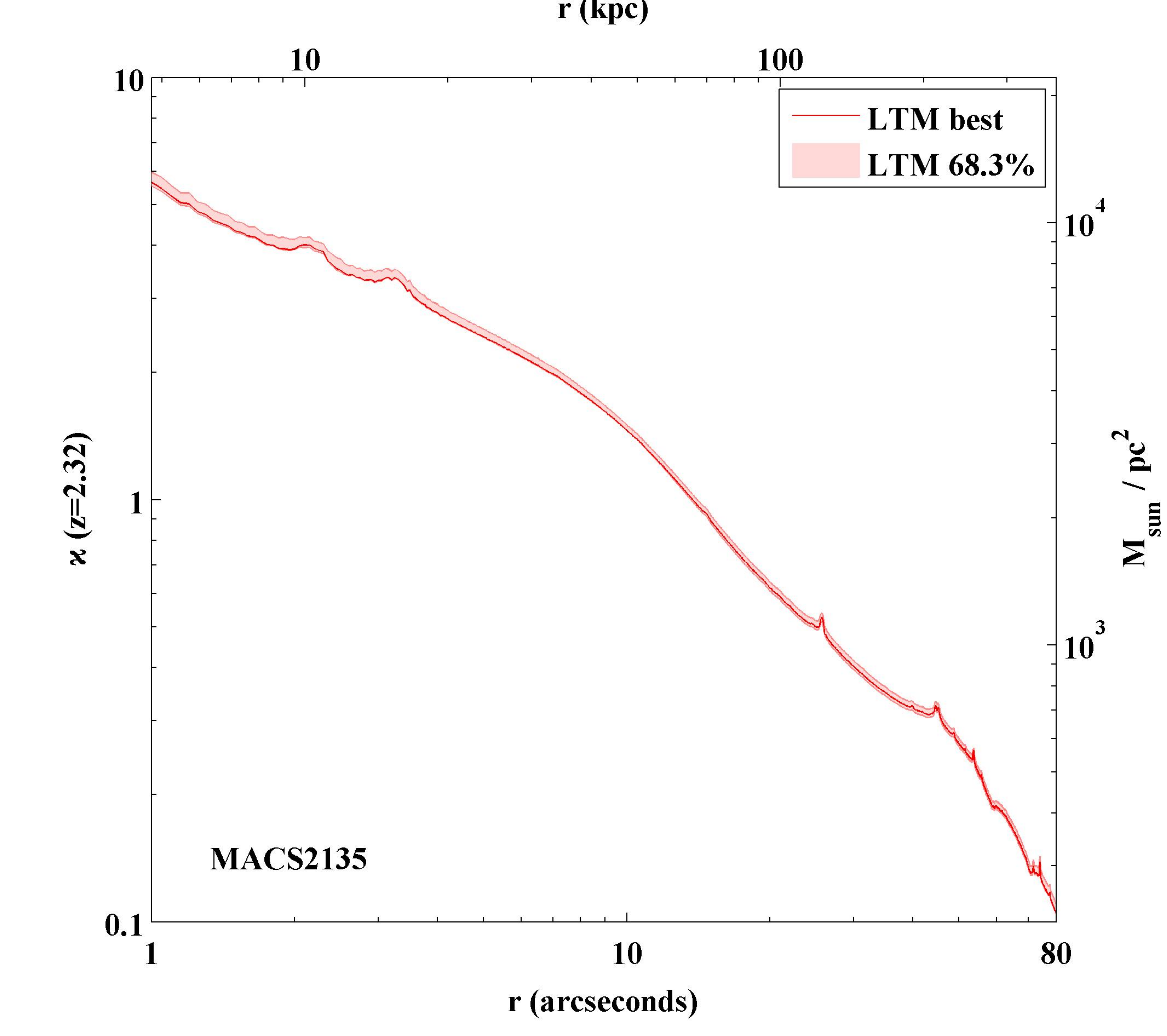}
 \end{center}
\caption{The resulting mass model for MACS2135. Upper subfigure shows the mass density, $kappa$ map for $z_{s}=2.32$, and the bottom subfigure shows the radially-averaged $kappa$ profile, with $1\sigma$ errors. }\vspace{0.5cm}
\label{fig2}
\end{figure}

\begin{deluxetable*}{ccccccc}
\tablewidth{0pc}
\tabletypesize{\footnotesize}
\tablecaption{\small{Multiple Images and Candidates} \label{multTable}}
\tablecolumns{6}
\tablehead{
\colhead{Arc ID} & \colhead{R.A} &\colhead{DEC.} &\colhead{$z_{phot}$ [95\% C.I.]} &\colhead{$z_{model}$ [95\% C.I.] } & \colhead{Comments}  \\ 
}   
\startdata
1.1 & 21:35:11.714 & -01:02:53.58 & -2.3259              & \nodata &  S10, I10 \\
1.2 & 21:35:11.613 & -01:02:52.35 & "                   & \nodata & "  \\
1.3 & 21:35:15.573 & -01:03:12.90 & ''                   & \nodata &  " \\
\hline
\rule{0pt}{3ex}1.11 & 21:35:11.796 & -01:02:54.45 & ''             &     \nodata &  " \\
1.21 & 21:35:11.521 & -01:02:50.98 &      ''          &   \nodata & "  \\
\hline
\rule{0pt}{3ex}2.1 & 21:35:11.580 & -01:03:34.79 &  -2.32     &    \nodata             &   S10  \\
2.2 & 21:35:12.195 & -01:03:37.19 &   ''       &    \nodata             & "   \\
2.3 & 21:35:10.040 & -01:03:18.66 &  ''       &    \nodata             & "  \\
\hline
\rule{0pt}{3ex}3.1 & 21:35:14.001 & -01:02:41.01 & 1.66 [1.40 1.92] &    2.00 [1.96 2.02]             & \nodata   \\
3.2 & 21:35:13.086 & 01:02:28.71 & 1.39 [1.16 1.62] &    "            & \nodata   \\
3.3 & 21:35:11.265 & -01:02:30.06 & 1.29 [1.07 1.51] &    "             & \nodata   \\
\hline
\rule{0pt}{3ex}4.1 & 21:35:11.762 & -01:02:52.46 &  \nodata  &  1.40 [1.40 1.51]             & \nodata             \\
4.2 & 21:35:11.724 & -01:02:51.56 & \nodata   &"            & \nodata             \\
c4.3 & 21:35:15.626 & -01:03:11.64 &  \nodata    &      " & \nodata            \\
\hline
\rule{0pt}{3ex}5.1 & 21:35:11.893 & -01:02:54.88 &  \nodata &      3.00 [2.94 3.13]  & \nodata            \\
5.2 & 21:35:11.584 & -01:02:48.73 & 1.82 [1.54 2.10]&       " & \nodata            \\
c5.3 & 21:35:15.630 & -01:03:12.88 &   \nodata &       " & \nodata            \\
\hline
\rule{0pt}{3ex}6.1 & 21:35:13.191 & -01:02:48.51 & 1.52 [1.27 1.77]&      1.14 [1.14 1.17]  & \nodata            \\
6.2 & 21:35:13.476 & -01:02:56.91 & 1.65 [1.39 1.91] &       " & \nodata          \\
c6.3 & 21:35:11.000 & -01:02:38.31 & 1.67 [1.41 1.93]  &"            & other candidates nearby \\
\hline
\rule{0pt}{3ex}7.1 & 21:35:13.039  & -01:02:44.57 &  1.81 [1.53 2.08]&      1.72 [1.71 1.77]  & \nodata   \\
7.2 & 21:35:13.713 & -01:03:10.25 & 2.23 [1.91 2.55] &      "  & \nodata       \\
7.3 & 21:35:12.192 & -01:03:14.31 &  \nodata &    "  &    \nodata          \\
7.4 & 21:35:10.015 & -01:02:48.26 & 1.53 [1.27 1.79]&       "  & \nodata            \\
\hline
\rule{0pt}{3ex}8.1 & 21:35:12.918 & -01:02:45.61 & 1.50 [1.25 1.75]&       1.97 [1.94 2.02] & \nodata        \\
8.2 & 21:35:13.581 & -01:03:15.51 & 1.64 [1.38 1.90]&        "  & \nodata          \\
8.3 & 21:35:12.195 & -01:03:16.55 &  \nodata &       " & \nodata      \\
8.4 & 21:35:09.940 & -01:02:50.88 & 1.65 [1.39 1.91]&      " & \nodata   \\
\hline
\enddata
\tablecomments{$\emph{Column 1:}$ arc ID . ``c'' stands for candidate where identification was more ambiguous and the image was not used as a constraint.\\
$\emph{Columns 2 \& 3:}$ RA and DEC in J2000.0. \\
$\emph{Column 4:}$ Photometric redshift and 95\% C.L. from BPZ. If a spectroscopic redshift is available it is marked with a minus sign, along with its references in the \emph{comments}.\\
 $\emph{Column 5:}$ Predicted and 95\% C.L. redshift by our lens model for systems lacking spectroscopic redshift.\\
 $\emph{Column 6:}$ Comments/References. S10 = \citet{Swinbank2010Natur}; I10=\citet{Ivison2010}.\\
Note that many of the model-predicted redshifts only marginally agree with the photometric redshifts within the errors.  This could be a result of a systematic bias in the lens model, as often revealed when comparing different mass modeling techniques \citep{Zitrin2014CLASH25}, and especially, given the lack of sufficient spectroscopic redshifts for multiple-images, which usually help increase the precision of mass models \citep[e.g.][]{Smith2009M1149, Kawamata2016modelsHFF,Grillo2015_0416,Jauzac2014M0416,Johnson2014HFFmodels}.}
\end{deluxetable*}

\section{Lens Model} \label{lensmodel}
We use the light-traces-mass (LTM) modeling technique by \citet[][see also \citealt{Broadhurst2005a}]{Zitrin2009_cl0024} to construct a lensing model for MACS2135. We briefly describe the method here and refer the reader to the said works for full details. Our model generally consists of three components: a galaxy component, which is a superposition of all galaxy mass contributions; a dark matter (DM) smooth component, which is a smoothed version of the galaxy component; and a two-parameter external shear. 

As mentioned in \S 2, we start with the list of red-sequence cluster galaxies and their photometry. To construct the galaxy component, each member galaxy is assigned with a power-law mass density distribution, scaled by its luminosity, where the superposition of all galaxies makes the total galaxy component of the model. The power-law exponent is the same for all galaxies and is a free parameter of the model. 

To obtain the DM smooth component, the galaxy component mass density map is then smoothed with a 2D Gaussian, whose width is a free parameter of the model. In that respect, both the galaxy and DM components follow the light distribution in an approximate sense as desired, since the finite statistical number of galaxies means we cannot 
expect an identical distribution. The two components are then combined with a relative weight, which is the third free parameter of the model. The overall normalization constitutes the fourth free parameter. A two component external shear is then added to allow for more flexibility, introducing effective ellipticity to the lens model. We also allow here for the mass of a few central bright cluster galaxies including the BCG, as well as their ellipticity and position-angle, to be freely optimized by the minimization procedure. 

In particular, after various iterations, the final model was minimized with seven key cluster members kept free: these include the three central bright galaxies embedded in the BCG's light in Fig. 1 (the relative weight between these affects the detailed shape of system 1's arc, for example), where also the ellipticity and PA of the two brighter galaxies were left to be freely optimized; a bright elliptical (21:35:15.192, -01:03:01.70) next to images 1.3/4.3/5.3, which are the farthest from the critical curves; a bright galaxy (21:35:11.193, -01:03:38.27) next to the most southern critical curve's tip and a bright galaxy (21:35:14.870, -01:02:19.53) above its northern tip, since these help refine the shape of the curves in those regions;  and another bright elliptical sitting within the critical curves (21:35:12.192, -01:03:21.44), to allow for somewhat more flexibility south of the BCG. We note that the choice of cluster members to be left free is somewhat subjective, guided by the lensed image distribution and proximity to individual member galaxies, and naturally there may be other valid permutations in this respect.

The best fit is then obtained through a long (few thousand step) Monte Carlo Markov Chain (MCMC), via a $\chi^{2}$ criteria minimizing the distance between the observed multiple images and those generated by the mass model. To infer the position of predicted images for each system, we use as source position the \emph{mean} source position obtained by delensing to the source plane the different images of this system. We then relens this source back to the image plane to predict the appearance of the multiple images.

One of the advantages of the LTM technique is that a well-guessed preliminary model can be constructed even with very few, or even none, multiple-images as input. In return the technique excels in predicting the location of multiple-images that can be then incorporated as constraints to iteratively improve the fit. We iteratively go over arclets and blue $z_{phot}\gtrsim1$ galaxies in the core of MACS2135 and delens-relens them with a preliminary LTM model to match their counter images in the data (based then, also, on a by-eye, SED and photo-$z$ examination, in addition to the model's prediction).  Similar to the success of this approach in other clusters \citep[e.g.][]{Broadhurst2005a,Zitrin2009_cl0024,Zitrin2013M0416,Zitrin2014CLASH25}, we identify here 6 new multiple images systems. After identifying the bulk of multiple images presented here in Table \ref{multTable} and Figure \ref{fig1}, we run our final mass reconstruction.  

We fix the redshifts of systems 1 and 2, that have spectroscopic redshifts of $z\simeq2.32$ from \citet{Swinbank2010Natur}, and allow for the redshift of all other systems to be optimized as free parameters in the minimization procedure, around their respective mean photo-$z$'s. With these, our final mass model has in total 23 free parameters. The resulting critical curves of our model, for $z=2.32$ are shown in Figure \ref{fig1}. The resulting mass density distribution, and mass profile, are shown in Figure \ref{fig2}.

Our final model has an image reproduction \emph{rms} of 1.68\arcsec, and a $\chi^2$ of 35.4 ($\chi^2/DOF =35.4/18\simeq2$), using a positional uncertainty of 1.4\arcsec. \citet{Zitrin2014CLASH25} found that while the true positional measurement error is small, this value is more representable of also systematic uncertainties between different modeling techniques, and folds within also discrepancies generated by random structures along the line of sight \citep[e.g.][]{Host2012LOS}. In practice, we use a posteriori the value of 1.4\arcsec\ for extracting the errors around the best fit, but the minimization itself was performed using a positional uncertainty of 0.5\arcsec\ for most images, and 0.25\arcsec\ for the radial systems 1, 4, and 5. We also used the parity of these three systems to force the radial critical curve to pass between the pairs of radial images (this was done by ``punishing'' the $\chi^2$ term if the input parity was violated). 

The final  \emph{rms} of our model, while reasonable and a common value for lensing analyses, is somewhat higher than that recently reported in different parametric modeling results, especially in regards to the thoroughly analyzed HFF clusters \citep[e.g.][]{Johnson2014HFFmodels,Grillo2015_0416,Jauzac2014M0416,Kawamata2016modelsHFF}. This is perhaps not surprising, since the LTM method was developed to yield maximal prediction power with a minimum of free parameters. Indeed, here we leave more galaxies to be freely weighted, as well as some of their ellipticities so the final number of free parameters is significantly increased. This is done to allow for somewhat more flexibility in the reproduction of images, but the overall solution still remains largely coupled to the light distribution. In other words, the LTM method may be less spatially flexible due to its coupling to the light distribution, than other purely analytic techniques with (often) inherently large numbers of free parameters, but has unprecedented prediction power to locate multiple images, even when none are known a priori and the number of free parameters is minimal (for example, \citealt{Zitrin2012UniversalRE} showed that a good, approximate solution can be obtained with as few as one parameter, the adopted mass-to-light relation). This prediction power has led us to constantly find, guided by preliminary LTM models, large numbers of multiple images previously undetected in many clusters. Additionally, on a technical note, it should be noted that the LTM model is in practice constructed on a grid, whose resolution is, for speed-up purposes, comparable to or somewhat lower than that of HST. In high magnification regions the round-up of the average source position to the grid's lower resolution pixel scale, introduces a finite, non-negligible \emph{rms} error of order 0.1\arcsec\ per system, contributing to the global imprecision of the model (but, importantly, without harming its reliability nor prediction power). These points have been recently emphasized in more length in a community effort to compare lens modeling techniques to simulated clusters \citep{Meneghetti2016SIMSCOMP}, and we refer the interested reader to that work for more discussion on this end.

We measure an effective Einstein radius of $\theta_{e}(z=2.32)=27\pm3\arcsec$ for the redshift of systems 1 and 2. This radius is the circular equivalent radius given the total enclosed area, i.e. $\sqrt{A/\pi}$. The critical curves for this redshift enclose $1.12 \pm0.16 \times10^{14}$ $M_{\odot}$.

\section{Discussion and Summary} \label{discussion}

Using HST images coupled with our LTM mass modeling we have identified, in addition to the two systems previously known \citep{Swinbank2010Natur},  six new multiply-imaged systems in MACS2135, roughly quadrupling the number of constraints to map the matter distribution in this cluster. We have correspondingly constructed a significantly improved mass model for MACS2135, which we present here and make available for the astronomical community. Our model agrees well with this cluster being a large lens, as perhaps is expected given the distance of system 2 from the BCG, and in broad agreement with the estimate presented in \citet{Swinbank2010Natur}. 

Only a small fraction of the clusters well-studied in the literature are known to exhibit Einstein radii exceeding $\gtrsim30\arcsec$ (nominally, for sources at redshifts around $z\sim2$). For example, only few out of the 25 X-ray selected CLASH clusters have Einstein radii comparable to, or slightly larger than that of MACS2135, and only a few clusters previously analyzed have considerably larger critical areas, e.g. Abell 1703, \citep{Limousin2008}; MACS 0717 \citep{Zitrin2009_macs0717}; RXJ1347 \citep{Zitrin2014CLASH25}, Abell 1689 \citep{Broadhurst2005a}; A370 \citep{Richard2010A370}; RCS2 J232727.6-020437, \citep{Sharon2015RSC2Big}; SDSS J120923.7+264047 \citep{Ofek2008Arc}; or CL0024 \citep{Zitrin2009_cl0024}. Indeed, thanks to their large critical areas all of these clusters show numerous multiply-imaged background galaxies, typically revealed in deep HST imaging. Additionally, the HFF clusters for example, aside for the giant lens MACS0717 \citep{Zitrin2009_macs0717} and perhaps A370 \citep{Richard2010A370}, show typically Einstein radii of $\sim25-30\arcsec$. Here we add to this important list MACS2135, showing that despite it current relatively shallow imaging, it also lenses an abundance of highly magnified, multiply-lensed background sources, and is comparable in size to the typical HFF cluster. 

Finding large and prominent lensing clusters is useful for probing the massive-end of the cluster mass function \citep{Zitrin2009_macs0717,Waizmann2012JeanClaude0717,Redlich2014}, for constraining cosmological models \citep{OguriBlandford2009}, and also for studying the DM, substructure, morphology and merging properties of the clusters \citep{Merten2011,Harvey2015Sci}. Large lenses also increase the chances for finding very high redshift galaxies often pushing the redshift limit  \citep[e.g.][]{Kneib2004z7,Coe2012highz} and in the case of multiple images we can use the separation between the images to provide a purely ``geometric" distance for the source as a means of testing the often ambiguous photometric redshift  \citep{Zitrin2014highz}. In fact, two high-$z$ candidates have already been reported in MACS 2135 \citep{Repp2016HighzMACS}, one of which our model predicts should lie nearly on top the critical curves for high redshift, and thus might be highly magnified and potentially multiply imaged. We leave further examination of this candidate for other, dedicated work. The lensing approach to studying high-$z$ galaxies is sensitive to the faint-end slope of the luminosity function, and complements the field work with Hubble that is also uncovering relatively luminous high-$z$ galaxies over wider areas \citep[e.g.][]{Ellis2013Highz,Bouwens2015LF10000}.

It should be appreciated that not only the Einstein radius of a lens is important in assessing the lensing efficiency of various clusters, but as we have shown before, other factors must be considered, such as the magnification distribution (which is related to the gradient of the central mass distribution), substructure and sub-clumps that add non-linearly to the magnification \citep{Redlich2014}, or the ellipticity of mass distribution which enhances the lensing cross-section \citep{Zitrin2013M0416}, as well as of course, the redshifts involved and the magnification bias which depends on the slope of the luminosity function \citep{Coe2014FF}.

We conclude that MACS2135 is amongst the top lenses currently known, especially in terms of it critical area, and will benefit from future attention. This includes deeper space imaging to uncover very distant high-redshift dropouts in the NIR, and as a compelling candidate target, in this respect, for JWST.

\vspace{0.01cm}

\section*{acknowledgments}
We thank the anonymous reviewer of this work for useful comments. Support for this work was provided by NASA through Hubble Fellowship grant \#HST-HF2-51334.001-A awarded by STScI, which is operated by the Association of Universities for Research in Astronomy, Inc. under NASA contract NAS~5-26555.This work is in part based on previous observations made with the NASA/ESA Hubble Space Telescope. 


\end{document}